\documentclass[%
 aip,
 amsmath,amssymb,
 twocolumn,
 floatfix,
 reprint,%
]{revtex4-2}

\usepackage{graphicx}
\usepackage{dcolumn}
\usepackage{bm}
\usepackage{xcolor}
\usepackage{url}

\usepackage{booktabs}
\usepackage{makecell}
\usepackage{enumitem}

\usepackage[T1]{fontenc}
\usepackage{textcomp,gensymb} 

\usepackage{hyperref} 
\hypersetup{
    colorlinks=true,
    linkcolor=blue,
    filecolor=magenta,      
    urlcolor=cyan,
}
\begin{document}


\title{An online readout CMOS detector to measure ion spectra produced by high-repetition-rate high-power lasers}

\author{K. Burdonov} 
\affiliation{LULI - CNRS, CEA, Sorbonne Universit\'e, Ecole Polytechnique, Institut Polytechnique de Paris - F-91128 Palaiseau cedex, France}
\affiliation{Sorbonne Universit\'e, Observatoire de Paris, Universit\'e PSL, CNRS, LERMA, F-75005, Paris, France}
\affiliation{IAP, Russian Academy of Sciences, 603155, Nizhny Novgorod, Russia}
\affiliation{JIHT, Russian Academy of Sciences, 125412, Moscow, Russia}	
\author{R. Lelievre} 
\affiliation{LULI - CNRS, CEA, Sorbonne Universit\'e, Ecole Polytechnique, Institut Polytechnique de Paris - F-91128 Palaiseau cedex, France}

\author{P. Forestier-Colleoni} 
\affiliation{Université Paris-Saclay, CEA, CNRS, LIDYL, 91191 Gif-sur-Yvette, France}

\author{T. Ceccotti}
\affiliation{Université Paris-Saclay, CEA, CNRS, LIDYL, 91191 Gif-sur-Yvette, France}

\author{M. Cuciuc}
\affiliation{”Horia Hulubei” National Institute for Physics and Nuclear Engineering, 30 Reactorului Street, RO-077125, Bucharest-Magurele, Romania}

\author{L. Lancia} 
\affiliation{LULI - CNRS, CEA, Sorbonne Universit\'e, Ecole Polytechnique, Institut Polytechnique de Paris - F-91128 Palaiseau cedex, France}

\author{W. Yao} 
\affiliation{LULI - CNRS, CEA, Sorbonne Universit\'e, Ecole Polytechnique, Institut Polytechnique de Paris - F-91128 Palaiseau cedex, France}
\affiliation{Sorbonne Universit\'e, Observatoire de Paris, Universit\'e PSL, CNRS, LERMA, F-75005, Paris, France}

\author{J. Fuchs} 
\affiliation{LULI - CNRS, CEA, Sorbonne Universit\'e, Ecole Polytechnique, Institut Polytechnique de Paris - F-91128 Palaiseau cedex, France}

\date{\today}

\begin{abstract}
We present the design and the absolute calibration of charged particle high-repetition-rate online readout CMOS system tailored for high-power laser experiments. This system equips a Thomson parabola (TP) spectrometer, used at the Apollon petawatt scale laser facility to measure the spectra of protons produced by high-intensity laser-target interactions. The RadEye1 CMOS matrixes array detectors are paired with a custom triggering system for image grabbing. This allows us to register the proton and ion signals remotely at a high-repetition-rate. The latter is presently of one shot/min, but the frame grabbing enables the system to be compatible with modern high-power lasers running e.g. at 10 Hz. We detail here the implementation, in the harsh electromagnetic environment of such interactions, of the system, and its absolute calibration of the RadEye CMOS matrices, which was performed for proton energies from 4 MeV to 20 MeV.
\end{abstract}

\maketitle

\section{\label{sec:intro}Introduction}

The constant progress of laser technologies in the field of high-peak power lasers has opened the way to deliver optical pulses to a target chamber for laser-plasma interaction investigations at unprecedentedly high repetition rate \cite{danson_2019}. For example, the BELLA laser runs at the power of 1 PW and at 1 Hz repetition rate \cite{7934119}. At a higher power level of 10 PW, Apollon and ELI-NP will soon run at one shot per minute \cite{Papadopoulos:19,lureau_2020}. Such increase in the repetition rate of the lasers needs to be accompanied by a similar capability of detectors used to  measure the secondary ionizing radiations produced in laser-plasma interactions. Indeed, up to now, in high-energy and high-power laser experiments, passive detectors, such as Image Plate (IP) \cite{doi:10.1063/1.4944863,doi:10.1063/1.5109783}, CR-39 \cite{Jeong2017} or Radiochromic film (RCF) \cite{bolton2014}, are still mostly used to detect the spectra of secondary ion beams. These detectors are not compatible with high repetition rates as they need to be replaced after each shot, which means breaking the vacuum of the target chamber, which  dramatically increases the time required between each shot. 

Efforts have been undertaken in the community to develop in-line detection systems. These, for example, read, using a CCD/CMOS camera equipped with an objective, from outside of the vacuum chamber the fluorescence of a scintillating screen or a micro-channel plate (MCP). Such systems are routinely used for the detection of hard photons\cite{Liang2022}, or of relativistic electrons at conventional and plasma-based electron accelerators \cite{doi:10.1063/1.3310275,RevModPhys.90.035002} and accelerated positively charged ions \cite{doi:10.1063/5.0101859,Dover2017}. However, the need to transfer the image from the detector surface to the reading sensor outside is inevitably accompanied by loss of emitted photon signal, decreasing the sensitivity of such systems. Thus now arises the need for the development of fast reusable detectors able to directly detect the incoming radiation or particles, in order to allow optimizing in real-time their production.

In this paper we present the design and calibration of a new online readout CMOS system tailored for high-power laser experiments. It equips a Thomson parabola (TP) spectrometer \cite{doi:10.1080/14786440208637024} geared for recording the spectra of protons and positively charged ions originating from the interaction of multi-PW laser pulses with solid targets at the Apollon facility \cite{doi:10.1063/5.0065138}. In the sub-PW regime that was presently commissioned, these ions are produced by target normal sheath acceleration (TNSA) \cite{doi:10.1063/1.1333697}. The composite detector installed on this spectrometer uses an array of highly sensitive unprotected RadEye1 CMOS active sensors\cite{Reinhardt2013,Lindner2018}, Thus, no need of image transfer is required, providing the maximum sensitivity of the detector. To provide synchronization between the RadEye's frame-grabber and the facility trigger, we developed a custom triggering device. The absolute calibration of the RadEye CMOS response to protons with energies in the range from 4 MeV to 20 MeV was performed using an absolutely calibrated \cite{doi:10.1063/1.4826084,Martin2022} image plate grid, placed on top of the TP spectrometer detector.

The design of the spectrometer and of the CMOS detection system installed on it will be first described in Section~\ref{sec:design},  followed by the description of the triggering system in Section ~\ref{sec:trigger}. The results of the spatial proton energy calibration of the detector and the absolute calibration response of the RadEye1 CMOS to protons will be presented in Section~\ref{sec:calib}. Finally, we will draw our conclusions in Section~\ref{sec:summary}.

\section{\label{sec:design}Thomson parabola design}

\begin{figure}[htp]
    \centering
    \includegraphics[width=0.5\textwidth]{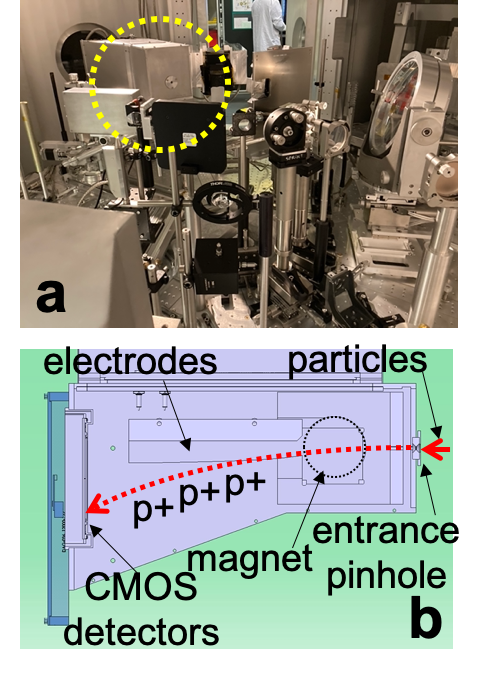}
    \caption{(a) Photography of the Thomson parabola (TP) spectrometer inside the SFA vacuum chamber of the Apollon facility (within the yellow dashed circle) (b) Schematic top view of the TP.}
    \label{fig:TP_scheme}
\end{figure}

The Thomson parabola spectrometer ANNA (ApolloN
thomsoN parabolA) was designed at CEA for the Apollon short focal area (SFA) vacuum chamber \cite{doi:10.1063/5.0065138} to be used in laser-plasma interaction experiments for the registration of the energy spectra of positively charged particles. A photo of the TP inside the chamber during an experimental campaign is shown in Figure~\ref{fig:TP_scheme}(a).

\begin{figure}[htp]
    \centering
    \includegraphics[width=0.4\textwidth]{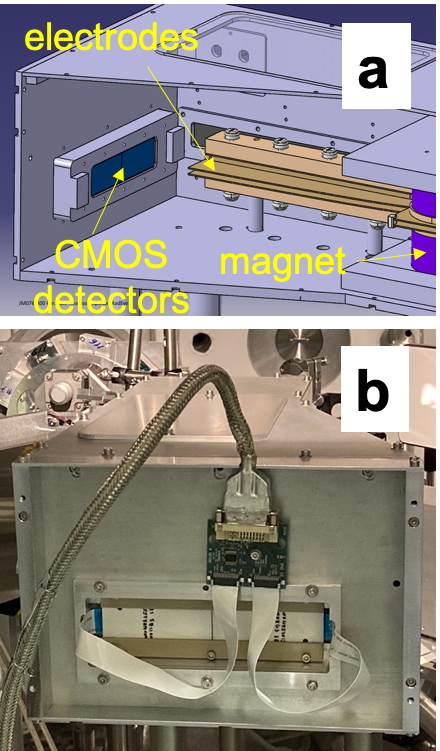}
    \caption{(a) A 3D image of the Thomson parabola (TP) spectrometer detector; (b) photograph of the mounting and cabling the CMOSes on the back of the TP assembly.}
    \label{fig:TP_detector}
\end{figure}

Figure~\ref{fig:TP_scheme}(b) presents the schematic top-view of the spectrometer. The diameter of the entrance pinhole may span between 50 µm and 2 mm. Two round magnets with 70 mm diameter produce 1.05 T magnetic field and two copper slabs with voltage difference up to 10 kV ($\pm 2$ kV were used for this particular campaign), serve to disperse in space the particles depending on their energy and mass-to-charge ratio\cite{doi:10.1080/14786440208637024} (see also Figure~\ref{fig:TP_detector}(a)). The spectrometer design allows to register proton spectra in the range from 1 MeV to 100 MeV.

The TP detector we developed is composed of two highly sensitive RadEye CMOS matrices that were mounted head to tail. A photo of the TP rear-end, showing the back of the detectors, is shown in Figure~\ref{fig:TP_detector}(b), demonstrating the mounting and cabling of the CMOSes inside the TP assembly.
The CMOS surface was impacted by the particles directly. To avoid electromagnetic perturbations induced onto the readout system by the harsh electromagnetic environment linked with the laser-target interactions (see Ref. \cite{doi:10.1063/5.0065138} for detailed measurements), the detectors were enclosed in a Faraday cage formed by the TP body and the cables were shielded (see Figure~\ref{fig:TP_detector}(b)). This proved very effective as the detectors were running unperturbed.

The CMOS detectors were triggered 0.4 s before the shot and readout right after. After each shot the signal was acquired remotely, allowing to use the spectrometer continuously over a series of shots without the need to access the detector, oppositely to, for example, commonly used imaging plates \cite{Lelasseux2020}, which requires the need to break the vacuum of the target chamber.

\begin{figure}[htp]
    \centering
    \includegraphics[width=0.45\textwidth]{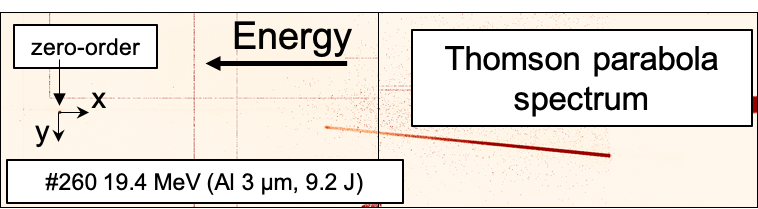}
    \caption{Typical spectral image obtained in a single shot with our detection system on the back of the Thomson parabola.}
    \label{fig:TP_data}
\end{figure}

A typical proton spectrum registered by the Thomson parabola spectrometer is shown in Figure~\ref{fig:TP_data}. A laser pulse with 9.2 J energy was focused onto the surface of a 3 $\mu$m aluminum target, providing a bunch of accelerated protons with a 19.4 MeV cutoff energy\cite{doi:10.1063/5.0065138}.
The whole data image consists of two separated images collected by the two RadEye CMOS matrixes. The so-called zero-order is the point-projection of the x-rays emitted by the target onto the detector. It serves as a reference from which the deflection of the charged particles can be accounted for.
A track of deflected protons is seen on the right-hand side of the image.  The proton track is parabolically curved, as is expected \cite{doi:10.1080/14786440208637024}.
The low energy part of the spectrum has been intentionally cut by a thick Al filter since it saturated the detector, as the number of particles increases exponentially toward the low energies\cite{Daido_2012}.
A  high-energy cut-off and a broad spectrum are clearly seen, which are both typical signatures of TNSA acceleration \cite{Daido_2012}.

\section{\label{sec:trigger}External triggering and synchronisation with facility}

The RadEye1 is a large area 512$\times$1024 pixel CMOS sensor that can be triggered by asserting its START signal. The associated Shad-o-Box readout electronics used in this experiment controls all the signals for the sensor based on an internal asynchronous frame timing. As this frame timing cannot be synchronized to an external source, a custom triggering system was developed that interfaced the incoming trigger signal from the laser to the control signals for the Shad-o-Box, ensuring that the sensor was not being cleared when the laser shot arrived and that the correct frame was read out.

\begin{figure}[htp]
    \centering
    \includegraphics[width=8.5 cm]{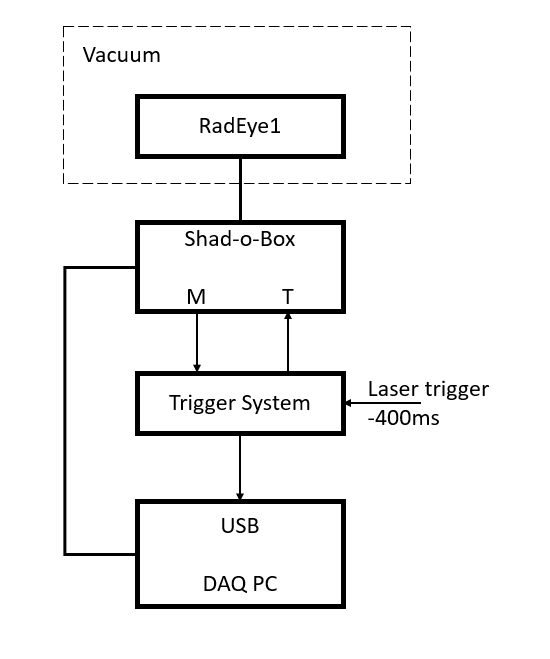}
    \caption{Integration of the trigger system in the data acquisition setup. The T line of the Shad-o-Box is asserted in order to inhibit both readout of the current frame and sensor reset, keeping the sensor in integration mode. The triggering system asserts this signal based on the laser trigger and the Shad-o-Box internal asynchronous frame timing, available on the M line. The USB connection to the PC signals the DAQ software that it should stop acquiring new frames to prevent overwriting of physics data.}
    \label{fig:trigger_schematic}
\end{figure}

The triggering system was built around a PIC16F18446 microcontroller that was programmed to accept triggers from the laser 400 ms in advance of the shot. A description of its interface to the rest of the experimental setup is available in Fig. \ref{fig:trigger_schematic}. Frames were acquired remotely after each shot, allowing us to use the spectrometer continuously over a series of shots without the need to access the detector by breaking the vacuum of the target chamber.

\section{\label{sec:calib}Absolute calibration of the system}

The absolute calibration of the system in energy and sensitivity was performed during the commissioning experiment session at the Apollon facility \cite{doi:10.1063/5.0065138}. The spectrometer was placed along the axis perpendicular to the target rear surface with its entrance pinhole having diameter D = 200 $\mu$m at the distance of L = 320 mm from the target to the pinhole, giving the spectrometer solid angle $\Delta\Omega_{TP} = \pi D^2/(4L^2)$ = 0.31 $\mu$sr.

\subsection{\label{sec:calib_energy_spatial}Energy calibration of the disperson onto the detector}

The first part of the spectrometer calibration procedure consisted in determining whether the real position of a proton of a certain energy on the detector corresponds to the expected position calculated on the basis of the calculated proton deflection in the magnetic field:

\begin{equation}
\label{eq:E_calib}
E [MeV] = \frac{(q \cdot B \cdot r (L + r/2))^2}{200 \cdot d[cm]^2 \cdot m_p \cdot q},
\end{equation}

where 
$m_p$ = $1.67 \times 10^{-27}$ kg is the proton mass, $q$ = $1.6 \times 10^{-19}$ C is the proton charge, $B$ = 1.045 T is the nominal magnetic field, $L$ = 0.252 m is the nominal distance between magnet and detector, $r$ = 0.07 m is the nominal magnetic field length, and $d$ is the distance between zero-order point on the detector and the point on the detector corresponding to a certain proton energy.

\begin{figure}[htp]
    \centering
    \includegraphics[width=0.45\textwidth]{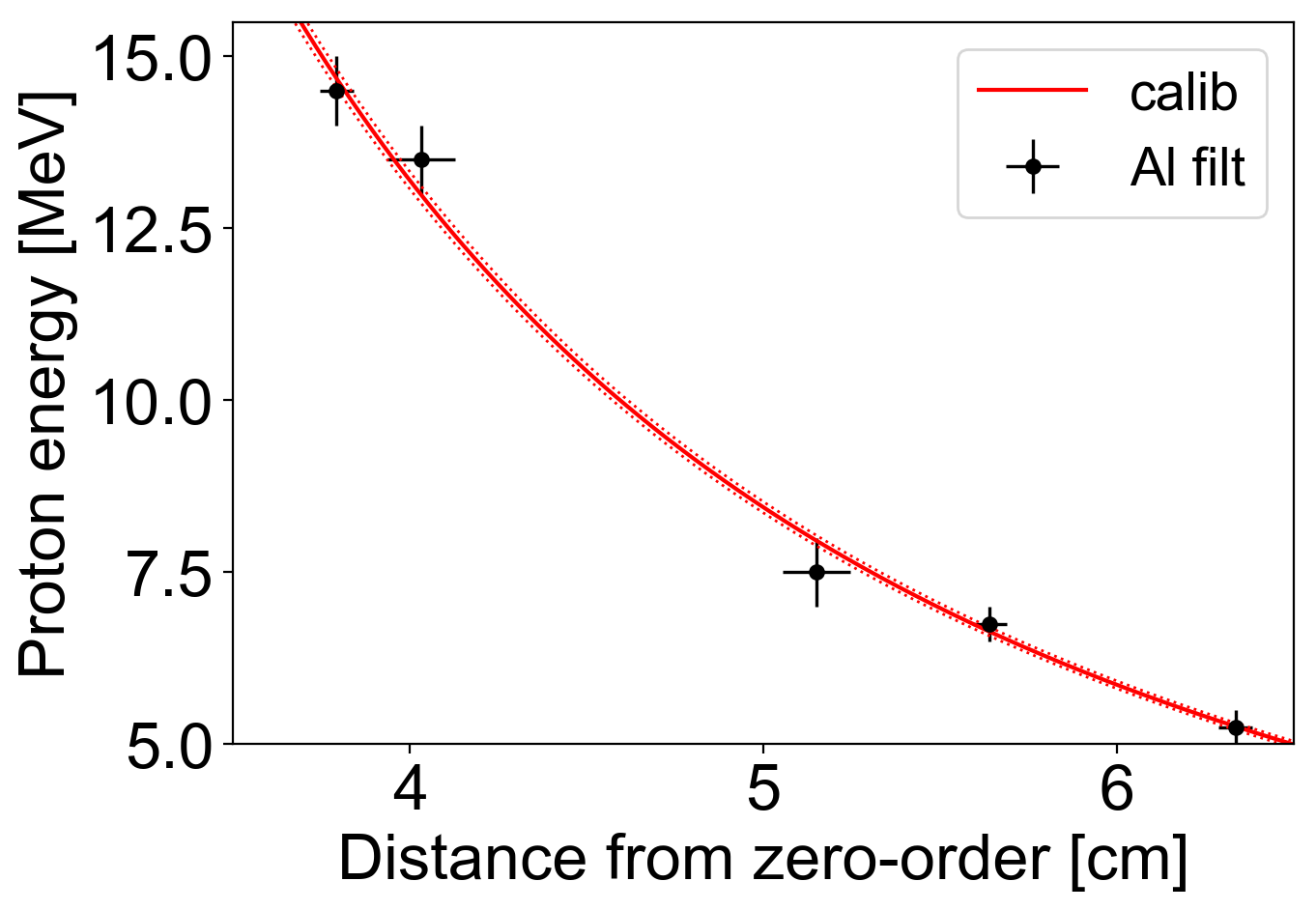}
    \caption{Proton energy dispersion calibration of the Thomson parabola.}
    \label{fig:TP_cal_spat}
\end{figure}

Aluminum filters with various known thicknesses were placed on top of the detector in order to block protons with the energy not high enough to go through the filters. The thickness of aluminum required to stop protons of a certain energy was calculated using the SRIM (The Stopping and Range of Ions in Matter) software \cite{ZIEGLER20101818} with a step of 0.5 MeV for proton energies in the range from 4 MeV to 7 MeV and with a step of 1 MeV in the range from 7 MeV to 18 MeV. Thus, an aluminum plate thickness of 0.2 mm can block proton with energy below an energy in the range of 5 MeV to 5.5 MeV; and similarly 0.3 mm, 0.4 mm, 1 mm and 1.2 mm thick plates can block protons with energies in the ranges of 6.5-7 MeV, 7-8 MeV, 13-14 MeV, and 14-15 MeV correspondingly. Distance from the zero-order point to the proton signal cut-off position can be obtained from the experimental data for each of mentioned above aluminum filters. As can be seen in the Figure \ref{fig:TP_cal_spat}, the red calibration curve representing the formula \ref{eq:E_calib} fits all the experimentally measured points (black dots).

\subsection{\label{sec:calib_abs}Absolute sensitivity calibration of the RadEye1 detector}

\begin{figure}[htp]
    \centering
    \includegraphics[width=0.45\textwidth]{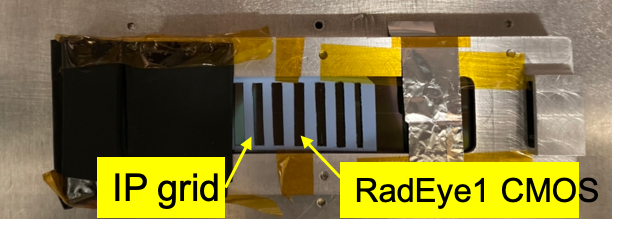}
    \caption{Photo of the IP grid that was positioned on top of the RadEye1 CMOS detector of the TP spectrometer for the absolute calibration measurement of the sensor sensitivity.}
    \label{fig:Detector_IP}
\end{figure}

To have the absolute calibration of the proton energy deposition on the CMOS, for one shot we placed a grid made of an image plate (IP)  with a known response to the proton impact \cite{doi:10.1063/1.2949388,Martin2022} in front the TP spectrometer detector, as shown in Figure \ref{fig:Detector_IP}). 

The IP was of the type BAS-TR 2025 (Fuji Photo Film Co. Ltd) with no protective layer, covering the detection phosphor, and it was scanned using the Fujifilm BAS-1800 II scanner.

The linear photostimulated luminescence (PSL), which is proportional to the number of incident protons, can be converted from the raw signal of the scanner on a logarithmic scale, namely Quantum level (QL), and which is stored in every single pixel, using the following formula:

\begin{equation}
\label{eq:QL_PSL}
IP_{PSL} = \Big(\frac{R}{100}\Big)^2 \times \frac{4000}{S} \times 10^{L \times \big(\frac{IP_{QL}}{2^G - 1} - \frac{1}{2}\big)},
\end{equation}


where $R$ = 50 is the scanner resolution (50 $\mu$m $\times$ 50 $\mu$m square pixel size); $L$ = 5 is the dynamic range (or Latitude); $G$ = 16 is the 16-bit depth; $S$ = 4000 is the sensitivity \cite{doi:10.1063/1.4893780}.

The equation that allows to retrieve the proton spectra, calculated using the IP scan, is:

\begin{equation}
\label{eq:IP_spectra}
\frac{dN(p^+)}{dE \cdot d\Omega} = \int_{-y}^y{IP_{PSL}\,dy} \cdot \frac{1}{dE_{IP} \cdot F_{IP} \cdot CAL_{IP} \cdot \Delta\Omega_{TP}},
\end{equation}

where $IP_{PSL}$ is the IP scanner data in PSL units (the integral along the $y$-axis corresponds to integrating all the signal perpendicularly to the direction of the energy dispersion); $dE$ is the proton energy dispersion; $F_{IP} = 73 \pm 3 \%$ (40 min after shot) is the IP fading with time from the exposure \cite{doi:10.1063/1.2949388}; $CAL_{IP}$ = $0.334 \cdot E [MeV] ^{-0.914}$ is the IP sensitivity to protons in the range of energies from 2.11 MeV to 20 MeV taken from \cite{doi:10.1063/1.2949388}. 

\begin{figure}[htp]
    \centering
    \includegraphics[width=0.45\textwidth]{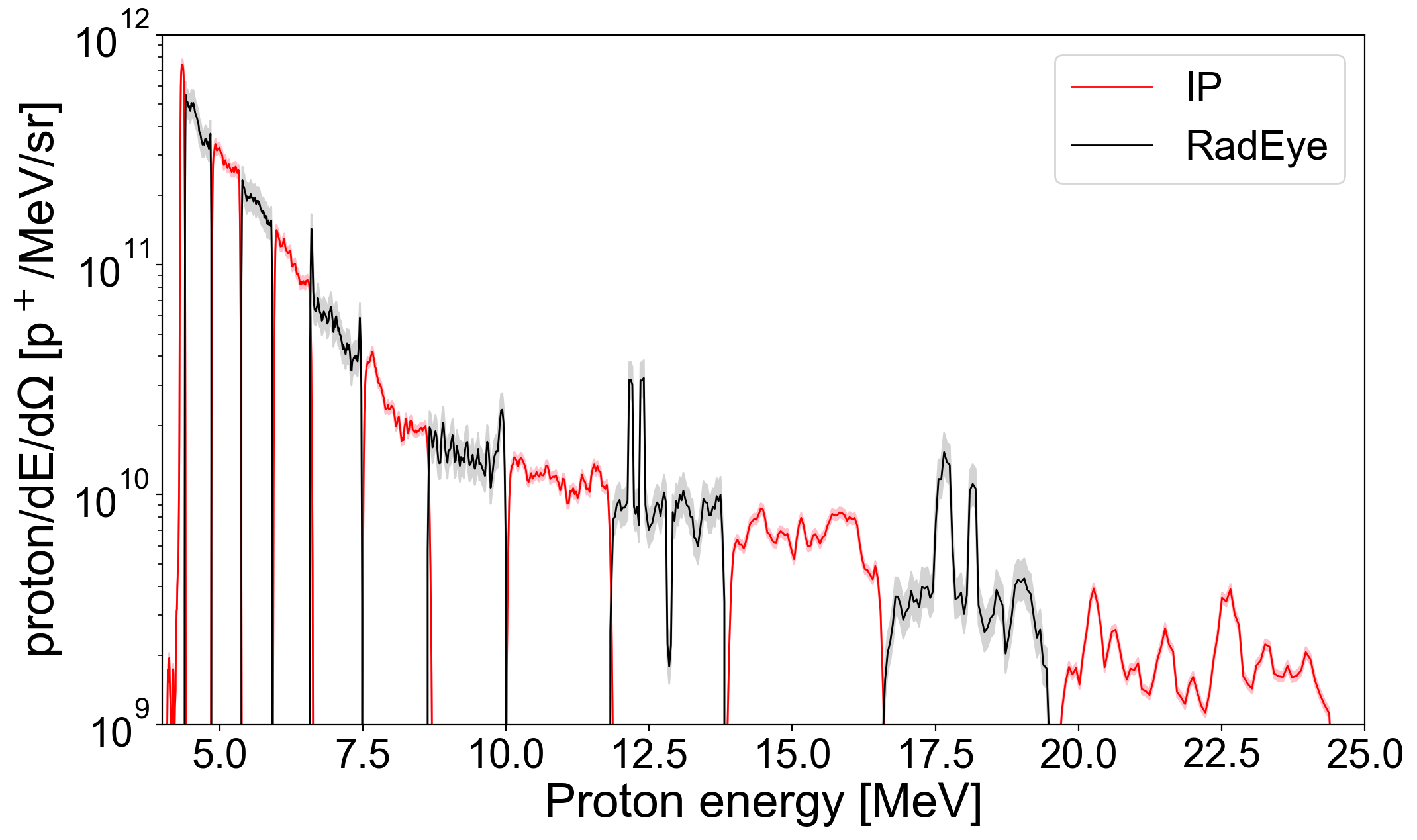}
    \caption{Calibration of the RadEye1 detector sensitivity by adjusting the spectrum that can be derived from it to the one derived from the slotted IP, and which absolute calibration is known.}
    \label{fig:TP_cal_abs}
\end{figure}

Finally, the equation to calculate the number of protons per MeV per steradians, basing on the data, registered by the RadEye1 CMOS array, is as follows:

\begin{equation}
\label{eq:RAD_spectra}
\frac{dN(p^+)}{dE \cdot d\Omega} = \int_{-y}^y{RE\,dy} \cdot \frac{1}{dE_{RE} \cdot \Delta\Omega_{TP} \cdot CAL_{RE}},
\end{equation}

where $RE$ is the raw data, registered by the CMOSes as TIFF format with 16 bit dynamic range; $dE$ is the proton energy dispersion; and 

\begin{equation}
\label{eq:RAD_calib}
CAL_{RE} = (10^3 \pm 50) \cdot E [MeV] ^{-1.5 \pm 0.5}
\end{equation}

is the desired function which allows to fit the absolute proton spectra, derived from the RadEye, to the one obtained with the help of the IP. This last calibration factor is obtained by a best fitting of the spectrum obtained from the RadEye to the one derived from the IP, as is demonstrated in Figure \ref{fig:TP_cal_abs}.

It should be mentioned that our fitting function differs from the one which was obtained with with the use of a CR39 detector for the proton energies in the range from 1 MeV to 3 MeV \cite{Reinhardt2013} and which can be written as $(1.09 \pm 0.12) \cdot$ E [keV].
Our additional power function coefficient of $-1.5$ can be neglected for the proton energies around 1 MeV, however for higher energies our more advanced fitting function has to be used.

\section{\label{sec:summary}Summary}

The design of a charged particle high-repetition-rate online readout CMOS system, tailored for high-power laser experiments, is presented.
The system is installed on the back of an ion spectrometer and consists in RadEye CMOS matrices that allow online recording of the measured proton spectra, even in the harsh electromagnetic environment induced by the high-intensity laser interactions. The calibration of the RadEye CMOS matrices was performed for  proton energies from 4 MeV to 20 MeV, giving the fitting function $CAL_{RE}$ = $(10^3 \pm 50) \cdot$E [MeV] $ ^{-1.5 \pm 0.5}$, allowing to restore the absolute energy spectrum of the proton signal based on the image recorded by the detector in TIFF format.

\begin{acknowledgments}
This work was supported by funding from the European Research Council (ERC) under the European Union’s Horizon 2020 research and innovation program (Grant Agreement No. 787539, project GENESIS), and by Grant ANR-17-CE30- 0026-Pinnacle from Agence Nationale de la Recherche. 
\end{acknowledgments}

\section*{Data Availability}

The data that support the findings of this study are available from the corresponding author upon reasonable request.

\section*{References}
\bibliographystyle{apsrev4-2-titles}
\bibliography{main}

\end{document}